%% file: MR_v2.tex
\documentclass[12pt]{article}
\pdfoutput=1
\usepackage{latexsym}\usepackage{epsfig,amssymb,euscript}
\usepackage{amsmath} \usepackage{bbm,slashed}
\usepackage[usenames,dvipsnames]{xcolor}
\usepackage{ulem}
\usepackage{cite}
\usepackage{cancel}
\usepackage{graphicx}
\usepackage{hyperref}

\newcommand{\bea}{\begin{eqnarray}}
\newcommand{\eea}{\end{eqnarray}}

\newcommand{\nn} {\nonumber}

\newcommand{\tx}{\text}

\newcommand{\bi}{\begin{itemize}}
\newcommand{\ei}{\end{itemize}}

\newcommand{\ben}{\begin{enumerate}}
\newcommand{\een}{\end{enumerate}}

\include{macros}
\catcode`@=12

\numberwithin{equation}{section}

\begin{document}

\begin{titlepage}

\begin{flushright}
SISSA 12/2016/FISI
\end{flushright}
\bigskip
\def\thefootnote{\fnsymbol{footnote}}

\begin{center}
\vskip -10pt
{\LARGE
{\bf
Scalar Multiplet Recombination  \\ \vskip 20 pt  at Large N and Holography
}
}
\end{center}

\bigskip
\begin{center}
{\large
Vladimir Bashmakov$^{1}$, Matteo Bertolini$^{1,2}$, \\ \vskip 10pt Lorenzo Di Pietro$^3$, Himanshu Raj$^1$}

\end{center}

\renewcommand{\thefootnote}{\arabic{footnote}}

\begin{center}
\vspace{0.2cm}
$^1$ {SISSA and INFN - 
Via Bonomea 265; I 34136 Trieste, Italy\\}
$^2$ {ICTP - 
Strada Costiera 11; I 34014 Trieste, Italy\\}
$^3$ {Weizmann Institute of Science - Rehovot 7610001, Israel}

\vskip 5pt

\end{center}

\noindent
\begin{center} {\bf Abstract} \end{center}
\noindent
We consider the coupling of a free scalar to a single-trace operator of a large $N$ CFT in $d$ dimensions. This is equivalent to a double-trace deformation coupling two primary operators of the CFT, in the limit when one of the two saturates the unitarity bound. At leading order, the RG-flow has a non-trivial fixed point where multiplets recombine.  We show this phenomenon in field theory, and provide the holographic dual description. Free scalars correspond to singleton representations of the AdS algebra. The double-trace interaction is mapped to a boundary condition mixing the singleton with the bulk field dual to the single-trace operator. In the IR, the singleton and the bulk scalar merge, providing just one long representation of the AdS algebra.

\vspace{1.6 cm}
\vfill

\end{titlepage}

\setcounter{footnote}{0}

%\tableofcontents

%%%%%%%%%%%%%%%%%%%%%%%%%%%%%%%%%%%%%%%%%%%%%
%%%%%%%%%%%%%%%%%%%%%%%%%%%%%%%%%%%%%%%%%%%%%%
\section{Introduction and Summary}\label{sec1}

Renormalization Group (RG) flow in Quantum Field Theory usually falls outside the regime of validity of perturbation theory. However, if an expansion parameter is available, like in the small-$\epsilon$ or the large-$N$ expansion, it may become possible to follow operators from the UV to the IR fixed point, and have direct access to interesting phenomena induced by the RG-flow. One such example is {\it multiplet recombination}: a primary operator that saturates the unitarity bound at the UV fixed point recombines with another primary operator, i.e. the latter flows to a descendant of the first at the IR fixed point, and the two distinct conformal families get mapped into a single one.

Recently, multiplet recombination was used to reproduce, via simple CFT arguments, perturbative calculations of anomalous dimensions in the $\epsilon$-expansion. This was first done for $O(N)$ scalar models in $4-\epsilon$ dimensions \cite{Rychkov:2015naa}, and later extended to the Gross-Neveu model in $2+\epsilon$ dimensions \cite{Ghosh:2015opa, Raju:2015fza}. In these examples, the short operator is a boson/fermion saturating the unitarity bound, which becomes long at the interacting fixed point due to its equations of motion. 

In this paper we will consider multiplet recombination in large-$N$ theories having a gravity dual description. Intuitively, holography should map multiplet recombination to a Higgs mechanism in the bulk. Indeed, when the protected operator is a conserved current that recombines due to a deformation that breaks the symmetry, the dual bulk gauge field is Higgsed and gets a mass. Here we will discuss the case in which the protected operator is a free scalar $\phi$, and couple it to a single-trace operator $O$ of the large-$N$ CFT via the interaction $\int d^d x\,\phi\,O$. We will see that also in this case there is a Higgs-like mechanism at work, albeit of a different kind, which exists only in AdS.

In order to study this problem, we find it useful to start considering two CFT single-trace operators $(O_1,O_2)$ of dimension $(\Delta_1,\Delta_2)$ with $\Delta_1 + \Delta_2 < d$, and thereafter take the decoupling limit $\Delta_1 \to \frac d 2 -1$. The relevant double-trace deformation 
\begin{equation}\label{deformation}
\int d^d x \, f \, O_1 O_2 ~,
\end{equation}
leads to an IR fixed point where $(O_1,O_2)$ are replaced by two operators $(\tilde{O}_1, \tilde{O}_2)$ of dimension $(d-\Delta_1, d-\Delta_2)$, respectively. In the limit $\Delta_1\to \frac d 2 -1$, many terms in the low energy limit of the correlators become analytical in the momentum, and can be removed by appropriate counterterms. Focusing on the physical part of the correlators, we will find that in this case multiplets recombine. In particular, $\tilde O_1 \propto \square \tilde O_2$, the IR dimensions being related as 
\be
\label{DrelIR}
\Delta^{IR}_1 = \Delta^{IR}_2 +2~,
\ee 
with $ \Delta^{IR}_2 = d - \Delta_2$. 

In the bulk the interaction \eqref{deformation} gets mapped into a non scale-invariant boundary condition for the scalar fields $(\Phi_1,\Phi_2)$ dual to $(O_1, O_2)$ \cite{Witten:2001ua, Berkooz:2002ug} (see also \cite{Mueck:2002gm, Gubser:2002zh, Gubser:2002vv, Hartman:2006dy, Papadimitriou:2007sj}). These bulk scalars are free at leading order in $1/N$ expansion. The presence of the coupling $f$ implies that the boundary modes of $\Phi_1$ and $\Phi_2$ get mixed. For $O_1$ and $O_2$ above the unitarity bound, the holographic analysis is standard, and the results agree with the field theory analysis. The limit $\Delta_1 \to \frac d2 -1$ should instead be treated with some care. One needs to rescale the field $\Phi_1$, otherwise the normalization of the two-point correlator of $O_1$ would vanish. Doing so, one sees that the on-shell action for $\Phi_1$ reduces to the action of a free scalar field living on the boundary of AdS, i.e. a singleton \cite{Flato:1980we, Fronsdal:1981gq, Duff:1998hj, Starinets:1998dt, Ohl:2012bk}. In the IR limit of the holographic RG-flow triggered by \eqref{deformation}, the singleton gets identified with a boundary mode of $\Phi_2$ corresponding to the VEV of the dual operator, i.e. the singleton becomes a long multiplet by eating-up the degrees of freedom of the bulk scalar.

The rest of the paper is organized as follows. In section 2 we perform the large-$N$ field theory analysis, and show that recombination takes place in the limit $\Delta_1 \to \frac d2 -1$. In section 3 we review the singleton limit in the bulk, and derive the holographic dual of the multiplet recombination flow. We conclude in section 4 with some comments on relations to previous work and possible future directions. An appendix contains the calculation of the variation of the quantity $\tilde{F}$ \cite{Giombi:2014xxa} induced by the flow \eqref{deformation}, which shows that $\delta \tilde F = \tilde F_{UV} - \tilde F_{IR} > 0$, in agreement with the generalized F-theorem advocated in  \cite{Giombi:2014xxa, Fei:2015oha}.

%%%%%%%%%%%%%%%%%%%%%%%%%%%%%%%%%%%%
%%%%%%%%%%%%%%%%%%%%%%%%%%%%%%%%%%%%
\section{Large-$N$ Multiplet Recombination: Field Theory}
\label{sec2}

Consider a free scalar $\phi$ coupled to a large-$N$ CFT through the interaction
\be
\label{def1}
\int d^d x f \phi \, O~,
\ee
where $O$ is a single-trace primary operator of dimension $\Delta < \frac d 2 + 1$, so that the deformation \eqref{def1} is relevant and triggers an RG-flow. 

At leading order in the large-$N$ expansion, one can integrate out the CFT sector and get the following non-local kinetic term for the scalar $\phi$
\be
\int d^d x \, f^2 \phi (-\square)^{\Delta - \frac d 2} \phi ~.
\ee
This term is dominant in the IR, indicating that $\phi$ flows to an operator of dimension $\Delta^{IR}_\phi = d - \Delta$.  In fact, the equation of motion for $\phi$ tells that in the IR $O = f^{-1} \square \phi$ becomes a descendant of $\phi$ with dimension $\Delta^{IR}_O = d - \Delta +2$. Therefore, in the IR $O$ disappears from the spectrum of primary operators, multiplets recombine, and the short multiplet of $\phi$ becomes long. 

As we will see, in order to understand the holographic dual phenomenon, it is useful to consider this flow as the limit of a double-trace flow induced by $f O_1 O_2$ when the dimension of $O_1$ saturates the unitarity bound. In the following subsections we review this double-trace flow and show that multiplet recombination emerges in the limit.

%%%%%%%%%%%%%%%%%%%
\subsection{Double-trace flow}\label{doubletraceflows}

Let us consider a large-$N$ CFT deformed by the double-trace interaction
\be
\label{pertf}
\int d^d x \, f \, O_1 O_2 ~,
\ee
where $(O_1, O_2)$ are single-trace primary operators of dimensions $(\Delta_1, \Delta_2)$, with $\frac{d}{2}-1 < \Delta_{1,2} < \frac d2$. Without loss of generality we will take $\Delta_1 < \Delta_2$ in what follows. One can conveniently analyze the perturbed CFT 
\be
\label{CFTd1}
S = S_{CFT} + \int d^d x \, f \, O_1 O_2 ~,
\ee
by introducing two Hubbard-Stratonovich auxiliary fields $\sigma_1$ and $\sigma_2$, and rewrite $S$ as
\be
\label{CFTd2}
S = S_{CFT} + \int d^d x \, \left( -f^{-1} \, \sigma_1 \sigma_2 + \sigma_1 O_1 + \sigma_2 O_2 \right)~.
\ee
Integrating $\sigma_1$ and $\sigma_2$ out gives the following relations
\be\label{integrateout}
\sigma_1 =  f O_2\,,\quad \sigma_2 =  f O_1~,
\ee
which, once substituted back into \eqref{CFTd2}, give the original action \eqref{CFTd1}. 

By performing the path integral in the CFT, one can derive an effective action for $\sigma_1$ and $\sigma_2$. To leading order at large $N$ all correlators of $O_1$ and $O_2$ factorize in a product of two-point functions. The resulting non-local effective action for the auxiliary fields is 
\bea
\label{effs1s2}
&& S[\sigma_1, \sigma_2] =  \\
&&- \int d^d x \left(f^{-1} \, \sigma_1(x) \sigma_2(x)  + \frac 12 \sigma_1(x) \int  \frac{d^d y}{(x-y)^{2\Delta_1}} \sigma_1(y) + \frac 12 \sigma_2(x) \int  \frac{d^d y}{(x-y)^{2\Delta_2}} \sigma_2(y)\right) \nn~.
\eea
Given that $\Delta_1$ and $\Delta_2$ are smaller than $\frac d 2$, the latter two terms dominate over the first, in the infrared. When only these terms are retained, $\sigma_1$ and $\sigma_2$ have IR correlators corresponding to operators with scaling dimension $d - \Delta_1$ and $d- \Delta_2$, respectively. 

Substituting \eqref{integrateout}, we hence obtain the following operators at the IR fixed point 
\begin{subequations}\label{IRdim}
\begin{align}\label{IRdim1}
\tilde{O}_1 = f O_2, \quad \Delta_1^{IR} = d- \Delta_1~,\\
\label{IRdim2}
\tilde{O}_2 = f O_1, \quad \Delta_2^{IR} = d - \Delta_2~.
\end{align}
\end{subequations}
The above result shows that the IR fixed point is the same as the one reached via the double-trace deformation $g_1O_1^2 + g_2 O_2^2$ \cite{Witten:2001ua}. This will be confirmed by the computation of the quantity $\tilde{F}$ \cite{Giombi:2014xxa} we do in the Appendix, where we show that the difference between the UV and IR values of $\tilde{F}$  induced by the flow \eqref{pertf} coincides with the one induced by the double-trace deformation $g_1 O_1^2+ g_2 O_2^2$.

%%%%%%%%%%%%%%%%%%%%%%%%%%%%%%%%%%%%%%%%%%%%%%%%%%%%%%
\subsection{Multiplet recombination}
\label{multrec}

We now take $\Delta_1 = \frac{d}{2} -1$, which means that $O_1$ decouples and becomes a free scalar, and consider again the perturbation \eqref{pertf} and the corresponding effective action \eqref{effs1s2}. The kernel of the non-local quadratic action for $\sigma_1$ is now $\frac{1}{(x-y)^{d-2}}$, which is the inverse of the Laplace operator. By the local change of variable
\be
\label{diagf}
\sigma'_1 = \sigma_1 - f^{-1} \square \sigma_2\,,\quad \sigma_2' = \sigma_2~,
\ee
one can cancel the mixing term in the action \eqref{effs1s2}, getting for the two-point function of $\sigma_1'$ just the contact term $\square \delta^d(x-y)$. Therefore, the following operator equation holds
\be
\label{mrel0}
\sigma_1' = 0 \Rightarrow \sigma_1 =  f^{-1} \square \sigma_2~. 
\ee
Using \eqref{integrateout}, \eqref{IRdim1} and \eqref{IRdim2}, we obtain the following operator relation at the IR fixed point
\be 
\label{mrrel1}
\tilde O_1 =  f^{-1} \square \tilde O_2~,
\ee 
signaling that multiplets recombine, i.e. $\tilde O_1$ becomes a descendant of $\tilde O_2$. Recall from eq.~\eqref{IRdim2} that $\tilde O_2 = f O_1$ has dimension $d-\Delta_2$ and, by \eqref{mrrel1}, $\tilde O_1 = f O_2$ has now dimension $d-\Delta_2 + 2$.

%%%%%%%%%%%%%%%%%%%%%%%%%%%%%%%%%%%
\subsection{A more general flow}

One might like to consider a more general double-trace deformation constructed out of $O_1$ and $O_2$, namely
\be
\label{moregeneral}
\int d^d x \left(  f \, O_1\,O_2 + \frac{g_1}{2} \, O_1^2 + \frac{g_2}{2} \, O_2^2\right)~,
\ee
and analyze the corresponding  RG-flow.\footnote{A similar quadratic interaction involving several single-trace operators was studied recently in the context of large-$N$ field theory in presence of disorder \cite{Aharony:2015aea}.} Introducing again Hubbard-Stratonovich auxiliary fields  one can recast the above action as
\be
\label{CFTd3}
S = S_{CFT} + \int d^d x \, \left[- \frac{1}{2 (f^2 - g_1 g_2)} \left( 2 f \sigma_1 \sigma_2  - g_2 \sigma_1^2 - g_1 \sigma_2^2 \right) + \sigma_1 O_1 + \sigma_2 O_2 \right]~.
\ee
Following the same steps as those of section \ref{doubletraceflows}, one ends-up with the following primary operators in the IR 
\begin{subequations}
\begin{align}
\tilde{O}_1 = g_1 O_1 + f O_2, \quad \Delta_1^{IR} = d- \Delta_1~,\\
\tilde{O}_2 = g_2 O_2 + f O_1, \quad \Delta_2^{IR} = d - \Delta_2~.
\end{align}
\end{subequations}
This shows that the IR fixed point is the same one reaches via the simpler deformation \eqref{pertf}, just the UV/IR operator dictionary is modified. So nothing qualitatively changes with respect to the previous analysis. 

Here again, one can safely take the decoupling limit $\Delta_1 \to \frac d2 -1$, getting a relation similar to \eqref{mrrel1}, the proportionality coefficient being now a function of $f, g_1$ and $g_2$
\be
\label{diagf12}
\tilde{O}_1 = \frac{f}{(f^2 - g_1 g_2)} \; \Box \tilde{O}_2~.
\ee 
So multiplets recombine also for this more general deformation. Notice that here the free operator is a massive one, its mass being proportional to $g_1$. Not suprisingly, for $f=0$ the deformation \eqref{moregeneral} does not trigger any multiplet recombination, as it is also clear from eq.~\eqref{diagf12}. The (massive) free operator simply gets integrated out, while $O_2$ flows to an operator of dimension $d-\Delta_2$.

Let us note that had we chosen $\frac d2 < \Delta_2 <d$, $O_2^2$ would have been an irrelevant deformation. This would not change much the story. Since one can always connect a CFT with $\Delta_2 > \frac d2$ to one with $\Delta_2 < \frac d2$  via an RG-flow with only $g_2$ turned on, there is no loss of generality in taking $g_2$ to be a relevant coupling, as we did from the outset.

As it is clear from eq.\eqref{CFTd3}, the hypersurface in the parameter space described by the equation $f^2 - g_1 g_2 =0$ needs a separate treatment. It is not difficult to see that in this case only one linear combination of the operators renormalizes, the IR dimensions of $\tilde O_1$ and $\tilde O_2$ being $d - \Delta_1$ and $\Delta_2$, respectively (the symmetry in the exchange $g_1 \leftrightarrow g_2$ is broken by the fact that we have chosen $\Delta_1 < \Delta_2$). This is a different IR fixed point with respect to previous cases. Actually, the same fixed point one reaches by deforming the CFT by $g_1$ only. Also in this special case one can take the decoupling limit, $\Delta_1 \to \frac d2 -1$. Proceeding the same way as before, one can see that the dimensions of $\tilde O_1$ and $\tilde O_2$ are now $\Delta_2 + 2$ and $\Delta_2$, respectively, indicating that multiplet recombination again holds. The IR fixed point is the same one reaches with a $g_1$ deformation only, which makes the free field disappearing from the IR spectrum, leaving only one primary of dimension $\Delta_2$. 

%%%%%%%%%%%%%%%%%%%%%%%%%%%%%%%%%%%
%%%%%%%%%%%%%%%%%%%%%%%%%%%%%%%%%%%
\section{Large-$N$ Multiplet Recombination: Holography}
\label{sec3}

In this section we will analyze the large-$N$ flows considered previously from a dual holographic perspective. Free operators of the QFT are dual to singleton representations of the AdS isometry group \cite{Flato:1980we,Fronsdal:1981gq} and some care is needed in dealing with them in the context of AdS/CFT. In particular, singletons do not enjoy any dynamics in the bulk. They correspond to propagating degrees of freedom only at the AdS boundary, and therefore the usual field/operator map should be properly interpreted. In what follows, we will first review how singletons can actually arise as a specific limit of ordinary bulk fields and how QFT correlators involving operators saturating the unitarity bound can then be consistently computed holographically using  ordinary AdS/CFT techniques. This will enable us to provide a holographic realization of QFT RG-flows enjoying multiplet recombination at large $N$.

%%%%%%%%%%%%%%%%%%%%%%%%%%%%
\subsection{Singleton Limit}

Consider a scalar $\Phi$ in $AdS_{d+1}$ with mass $m^2 = \Delta (\Delta -d)$, and $\Delta = \frac d2 -1 +\eta$. Eventually, we will be interested in the limit $\eta \to 0$. To leading order at large $N$ the scalar is free, and solving the Klein-Gordon equation we have the leading modes at the boundary
\be
\Phi(z, x) \underset{z\to 0}{\sim} (\Phi^-(x) z^{\Delta} + \Phi^+(x) z^{d-\Delta})(1+ O(z^2))~, 
\ee
where $z$ is the radial coordinate that vanishes at the boundary and $x \in \mathbb{R}^d$. Since $\Delta < \frac d 2$, the correct boundary condition is that $(d-2\D)\Phi^+(x) $ is fixed to coincide with the source of the operator of the boundary theory: $J(x)\equiv (d-2\D)\Phi^+(x)$. This implies, in turn, that one needs to include an additional boundary term to ensure that the bulk action is stationary \cite{Klebanov:1999tb}. After this is done, the renormalized on-shell action consists of the following boundary term in momentum space 
\be
S_{\text{on-shell}}^{\text{ren}} = \frac 12 \int_{z=0} \frac{d^d k}{(2\pi)^d}  \Phi^-[J(k)] J(-k)~.
\ee
The solution to the Klein-Gordon equation with the prescribed boundary condition and regular for $z\to \infty$ is
\begin{align}\label{bulksolution}
\Phi(k,z)_{\text{on-shell}} &  = -\frac { 1}{ \Gamma(1-\frac d 2 + \Delta)} \(\frac k2\)^{\Delta-\frac d2}J(k) z^{\frac d 2} K_{\frac d 2 - \Delta} (k z) \\ & \underset{\eta \to 0}{\sim}  -2 \eta~ k^{-1} J(k)z^{\frac d 2} K_1 (k z)~, \nonumber
\end{align}
where $K_{\frac d2 -\D}(kz)$ is the modified Bessel's function of the second kind. From the form of the solution we see that
\begin{align}
\Phi^-[J(k)] & = -\frac12 \frac{\Gamma( \frac d2 - \Delta)}{\Gamma(1-\frac d2 + \Delta) }\left(\frac k 2\right)^{2\Delta - d} J(k)\\ \nonumber & \underset{\eta \to 0}{\sim}- \frac{\eta}{2} \left(\frac k 2\right)^{-2} J(k)~.
\end{align}
Recalling that the two-point function is minus the second derivative of the effective action with respect to the source, we find that
\be
\langle O(k) O(-k) \rangle = \frac12 \frac{\Gamma( \frac d2 - \Delta)}{\Gamma(1-\frac d2 + \Delta) }\left(\frac k 2\right)^{2\Delta - d}\underset{\eta \to 0}{\sim} \frac{2\eta}{k^2} ~.
\ee
This shows that in order to get a finite result in the limit $\eta \to 0$ we need to rescale the source $J(x)$ of the operator as $J(x) = \frac{1}{ \sqrt{2 \eta}}\hat{J}(x)$, with $\hat{J}(x)$ finite in the limit. In terms of the bulk scalar field, this amounts to rescaling $\Phi(x, z) = \sqrt{2 \eta } \, \hat{\Phi}(x,z)$ with $\hat{\Phi}$ kept fixed. In this limit, the solution \eqref{bulksolution} goes to zero everywhere in the bulk, while the boundary term stays finite and becomes
\be
S_{\text{on-shell}}^{\text{ren}} \underset{\eta \to 0}{\rightarrow} \frac 12 \int_{z=0} \frac{d^d k}{(2\pi)^d} \hat{J}(k)\frac{1}{k^2} \hat{J}(-k)~.
\ee
This is the generating functional of a free scalar operator living on the boundary. Note that for $\eta \to 0$ we get $\hat{\Phi}^- = - k^{-2} \hat{J}(k)$. We can identify the free scalar operator $\phi$ on the boundary as $\phi \equiv \hat{\Phi}^-$. In fact, if we Legendre-transform back from $\hat{J}$ to $\phi$ the boundary term becomes $\frac 12 \int_{z=0} \frac{d^d k}{(2\pi)^d} \phi(k)k^2 \phi(-k)$, i.e. the action of a free scalar.

%%%%%%%%%%%%%%%%%%%%%%%%%%%%%%%%%%%%%%
\subsection{Holographic Recombination Flow}

We have now all ingredients to provide the holographic description of the large-$N$ flows discussed in section \ref{sec2}. We start considering two primary operators of the CFT with dimensions $\Delta_{1,2} < \frac{d}{2}$. The CFT operators are dual to two scalar bulk fields $\Phi_1 , \Phi_2$ having the following near boundary expansions
\begin{subequations}\label{bfields12}
\begin{align}
\Phi_1(z, x) \underset{z\to 0}{\sim} (\Phi^-_1(x) z^{\Delta_1} + \Phi^+_1(x) z^{d-\Delta_1})(1+ O(z^2)) ~,\\
\Phi_2(z, x) \underset{z\to 0}{\sim} (\Phi^-_2(x) z^{\Delta_2} + \Phi^+_2(x) z^{d-\Delta_2})(1+ O(z^2))~.
\end{align}
\end{subequations}
The deformation \eqref{pertf} is implemented by imposing the boundary condition\cite{Witten:2001ua}
\begin{subequations}\label{bdycond}
\begin{align}
J_1 \equiv (d-2\D_1)\Phi^+_1  + f  \Phi^-_2~,\\
J_2 \equiv (d-2\D_2)\Phi^+_2 + f  \Phi^-_1~,
\end{align}
\end{subequations}
where $J_{1}$ and $J_2$ are the sources for the field theory operators $O_{1}$ and $O_2$, respectively.

The solutions which are regular in the interior and have subleading boundary modes $\Phi^+_{1,2}$ are
\begin{subequations}\label{bulksolutiondeformation}
\begin{align}\label{bulksolutiondeformation1}
\Phi_1(k,z)_{\text{on-shell}} &  = -N_{\Delta_1}k^{\Delta_1- \frac d2} (d-2\D_1)\Phi^+_1z^{\frac d 2} K_{\frac d 2 - \Delta_1} (k z)~, \\
\label{bulksolutiondeformation2}
\Phi_2(k,z)_{\text{on-shell}} &  = - N_{\Delta_2} k^{\Delta_2- \frac d2}(d-2\D_2)\Phi^+_2 z^{\frac d 2} K_{\frac d 2 - \Delta_2} (k z) ~,
\end{align}
\end{subequations}
where
\be
N_\Delta = \frac { 2^{ \frac d2 -\Delta} }{\Gamma(1-\frac d 2 + \Delta)}~. 
\ee
From the explicit expressions \eqref{bulksolutiondeformation}, we can read-off the coefficients $\Phi^-_{1,2}$, and obtain a linear relation between $\Phi^+_{1,2}$ and $\Phi^-_{1,2}$. We can plug this in \eqref{bdycond} and solve for $(\Phi^-_1, \Phi^-_2)$ as linear functions of $(J_1,J_2)$. The solution is
\begin{align}\label{vevs}
\F_1^-[J_1,J_2]& =\frac{J_1 -fJ_2G_2}{1-f^2G_1G_2}G_1,~~~~\F_2^-[J_1,J_2] =\frac{J_2 -fJ_1G_1}{1-f^2G_1G_2}G_2,
\end{align}
where
\be
G_i(k)=-\frac12\frac{\G(\frac d2-\D_i)}{\G(1-\frac d2+\D_i)}\(\frac{k}{2}\)^{2\D_i-d}.
\ee
Using standard techniques, one gets the following renormalized on-shell boundary action consistent with boundary conditions \eqref{bdycond} 
\begin{align}
S_{\text{on-shell}}^{\text{ren}} &=\frac12 \int \frac{d^dk}{(2\p)^d} ~\((d-2\D_1)\Phi^+_1\Phi^-_1+(d-2\D_2)\Phi^+_2 \Phi^-_2 + 2f \Phi^-_1\Phi^-_2\)~.
\end{align}
Using \eqref{vevs} this can be rewritten in terms of the sources as follows 
\begin{align}\label{holgenfun}
S_{\text{on-shell}}^{\text{ren}}[J_1,J_2]=\frac12\int \frac{d^d k}{(2\pi)^d}~\bigg(&J_1(k)\frac{G_1}{1-f^2G_1G_2}J_1(-k)+J_2(k)\frac{G_2}{1-f^2G_1G_2}J_2(-k)\NO\\
-&2 J_1(k)\frac{f G_1G_2}{1-f^2G_1G_2}J_2(-k)\bigg)~.
\end{align}
This expression is equivalent to the the field theory result \eqref{effs1s2}. In order to see this, one needs to add the following local term to the above generating functional
\begin{align}\label{locterm}
S_{\text{local}}=-\int \frac{d^d k}{(2\pi)^d}~J_1(k)\frac{1}{f}J_2(-k)~,
\end{align}
and Legendre-transform. Identifying the Legendre-transformed fields with $(\frac1f \sigma_2, \frac1f \sigma_1)$ one gets precisely the Fourier transform of \eqref{effs1s2}, provided we identify the field theory coupling defined in section \ref{sec2} and the holographic coupling in the following manner
\be\label{couplingrel}
f^2_{\tx{hol}}=4\p^d\frac{\G(1-\frac d2 +\D_1)\G(1-\frac d2 +\D_2)}{\G(\D_1)\G(\D_2)}f^2_{\tx{ft}}~,
\ee
and pick the negative root for $f_{\tx{hol}}$ (it is generic in AdS/CFT that field theory couplings differ from holographic ones by such overall normalizations). After these identifications, one can repeat the analysis of section \ref{doubletraceflows} and obtain eqs.~\eqref{IRdim}.

In order to describe the phenomenon of multiplet recombination holographically, one has just to repeat the above analysis taking the singleton limit on the field $\Phi_1$, first. One should hence set $\Delta_1 = \frac d2 -1 + \eta $, rescale the source of $O_1$ as $J_1 = \frac{1}{\sqrt{2\eta}} \hat{J}_1$, rescale also the coupling as $f = \frac{1}{\sqrt{2\eta}} \hat{f} $, and eventually take the limit $\eta \to 0$, with the hatted quantities kept fixed. Doing so, and repeating previous steps one gets, eventually, equation \eqref{mrrel1}. Below, we find it instructive to adopt yet another (but equivalent) point of view. Instead of working with the effective action for $(\s_1, \s_2)$ we will work with the generating functional \eqref{holgenfun} itself. After the singleton limit the on-shell action, analogous to \eqref{holgenfun}, reads
\begin{align}
\label{holgenfunsing}
S_{\text{on-shell}}^{\text{ren}}=-\frac12\int \frac{d^d k}{(2\pi)^d}~\bigg(&\hat{J}_1(k)\frac{-k^{-2} }{1+\hat{f}^2k^{-2}G_2}\hat{J}_1(-k)+J_2(k)\frac{G_2 }{1+\hat{f}^2k^{-2}G_2}J_2(-k)\NO\\
+&2 \hat{J}_1(k)\frac{\hat{f} k^{-2}G_2}{1+\hat{f}^2k^{-2}G_2}J_2(-k)\bigg)~.
\end{align}
This action can be recast in the following way
\begin{align}
\label{holgenfunsing1}
S_{\text{on-shell}}^{\text{ren}}=\frac12\int \frac{d^d k}{(2\pi)^d}~\big((\hat{J}_1(k)+\frac{k^2}{\hat f}J_2(k)\big)\frac{k^{-2}}{1+\hat{f}^2k^{-2}G_2}\big(\hat{J}_1(-k)+\frac{k^2}{\hat f}J_2(-k)\big)~,
\end{align}
where certain contact terms have been dropped. We see that we are left with just one effective source $J_{eff}(x)=\hat{J}_1(x)-\frac{1}{\hat f}\Box J_2(x)$. Equivalently, the VEVs are related as
\begin{align}
k^2 \, \frac{\delta S_{\text{on-shell}}^{\text{ren}}}{\delta \hat{J}_1} = \hat{f} \; \frac{\delta S_{\text{on-shell}}^{\text{ren}}}{\delta J_2} ~.
\end{align}
This equation shows that, as a result of the interaction $f$, the VEV mode of the bulk scalar gets identified with the singleton $\hat{\Phi}^-_1$, and its original VEV mode is now obtained by applying $\square$ to $\hat{\Phi}^-_1$. This is the hallmark signature of multiplet recombination. Notice, finally, that in the IR  the behavior of the two-point function of the leftover primary operator is 
$\la \tilde{O}_2\tilde{O}_2\ra\propto k^{d-2\Delta_2}$, implying that at the IR fixed point we have a primary of dimension $\Delta^{IR}=d-\Delta_2$, in agreement with field theory analysis.

Summarizing, when $f=0$ there are two independent modes, i.e. the singleton $\hat \F^-_1$, which is just a boundary degree of freedom, and an ordinary bulk scalar, $\F_2$. They are associated to two independent sources, $\hat{J}_1(k)$ and $J_2(k)$. In contrast, at the IR AdS point, there exists only one independent source, $\hat{J}_1(k)- \frac{1}{\hat f} \Box J_2(k)$ and in turn only one scalar. $\F_2$  and the singleton merge into one bulk scalar whose VEV mode is $\hat \F^-_1$.

%%%%%%%%%%%%%%%%%%%%%%%%%%%%%%%%%%%
%%%%%%%%%%%%%%%%%%%%%%%%%%%%%%%%%%%

\section{Comments}
\label{sec4}
In this paper we have described multiplet recombination induced by coupling a large-$N$ CFT in $d$ dimensions to a free sector. Working at leading order in $1/N$, we have described  this phenomenon in field theory, and provided the holographic dual description. Let us comment on the relation with previous work on multiplet recombination in holography, and indicate some possible future directions.

Multiplet recombination in AdS$_5$/CFT$_4$ was studied in \cite{Bianchi:2003wx, Beisert:2004di, Bianchi:2004ww,Bianchi:2005ze}. In that case the recombination is not due to an RG-flow, rather it occurs as one moves away from the free point $g_{YM} = 0$ of $\mathcal{N}=4$ SYM on the line of the marginal coupling, and the higher-spin currents of the free theory get broken. Another instance of higher-spin multiplet recombination is the case of $O(N)$ vector models in AdS$_4$/CFT$_3$ \cite{Klebanov:2002ja, Girardello:2002pp,Leigh:2012mz}. The holographic dual description consists of a Higgs mechanism for the higher-spin gauge fields dual to the higher-spin currents of the free theory. The Higgs mechanism happens at tree-level in the example of $\mathcal{N}=4$ SYM, while it is a $1/N$ effect for the $O(N)$ vector models. 

The crucial difference between these examples and our setting is that in these examples one starts with $N\gg1$ free fields with a singlet condition, while we are considering only one free field. For this reason, in our setting there are no higher-spin gauge fields in the bulk. We only have higher spin currents associated to the singleton and supported on the boundary, and those are broken by the boundary condition.

A natural follow-up of our work would be to consider fermionic operators, along the lines of \cite{Heslop:2005ff,Laia:2011wf}, and study the analogous singleton limit and recombination in the bulk due to the boundary condition. 

The idea of multiplet recombination has been applied extensively in the literature in various contexts, to compute anomalous dimensions \cite{Anselmi:1998ms, Belitsky:2007jp, Rychkov:2015naa, Basu:2015gpa, Ghosh:2015opa, Raju:2015fza, Skvortsov:2015pea, Giombi:2016hkj}, to constrain the form of three-point functions \cite{Maldacena:2012sf}, or to find exactly marginal deformations \cite{Green:2010da}. In these examples one works perturbatively in a small parameter that controls the breaking of the shortening condition. In the case we consider, instead, the recombination happens at leading order in $1/N$, so we cannot apply these methods to obtain more information about the IR fixed point. It would be interesting to consider a set-up in which the shortening is violated by a multi-trace operator with a suppressed coupling at large $N$, as would follow for instance from an interaction $\int d^d x \phi \, O^2$, and see if similar techniques could instead be used in that case. Another open problem is to try to use multiplet recombination to compute anomalous dimensions in the IR fixed point of QED in $d=4-2\epsilon$ \cite{DiPietro:2015taa, Giombi:2015haa}.

We hope to report on some of these issues in the future.

%
%%%%%%%%%%%%%%%%%%%%%%%%%%%%%%%%%%%%%%%%%%%%%%%
\section*{Acknowledgements}

We are grateful to Bruno Lima de Souza, Ioannis Papadimitriou and Marco Serone for helpful discussions, and to Riccardo Argurio and Zohar Komargodski for useful comments on the draft. L.D.P. acknowledges support by the Israel Science Foundation center for excellence grant (grant no. 1989/14), the Minerva foundation with funding from the Federal German Ministry for Education and Research, by the I-CORE program of the Planning and Budgeting Committee and the Israel Science Foundation (grant number 1937/12), the ISF within the ISF-UGC joint research program framework (grant no. 1200/14) and the United States-Israel Binational Science Foundation (BSF) under grant number 2010/629. Support from the MPNS-COST Action MP1210 "The String Theory Universe" is also acknowledged.

%%%%%%%%%%%%%%%%%%%%%%%%%%%%%%%%%%%%%%%%%%%%%%%%%%%%%%
%

\appendix

%%%%%%%%%%%%%%%%%%%%%%%%

\section{Calculation of $\delta \tilde{F}$ for the double-trace flow $f O_1 O_2$}

In the following, we compute the leading large-$N$ variation of $\tilde{F}$ induced by the flow \eqref{deformation}. The quantity $\tilde F$ can be defined in a CFT where the dimension $d$ is promoted to a continuous parameter, and interpolates between the sphere free energy in odd dimensions and the $a$ anomaly in even dimensions \cite{Giombi:2014xxa}. In \cite{Giombi:2014xxa} several examples were provided for which $\tilde{F}$ decreases towards the IR, suggesting a generalization of the $a$- and F-theorems to continuous dimensions. In particular, it was proven that the generalized F-theorem holds for double-trace deformations. Here we follow the methods of \cite{Gubser:2002vv, Diaz:2007an, Giombi:2014xxa}.
 
The quantity of interest is defined as $\tilde{F} =  \sin\left(\frac{\pi d}{2}\right)\log Z^{\mathbb{S}^d}$, where $Z^{\mathbb{S}^d}$ is the partition function on a $d$-dimensional sphere. At leading order at large-$N$ the sphere partition function depends on the deformation $f O_1 O_2$ as 
\begin{equation}\label{pfsphere}
Z^{\mathbb{S}^d}_f = Z^{\mathbb{S}^d}_0\times \frac{1}{\sqrt{\det(1^{\mathbb{S}^d} - f^2 G^{\mathbb{S}^d}_1\star G^{\mathbb{S}^d}_2)}}~.
\end{equation}
This can be derived from the equivalent of action \eqref{effs1s2} for the theory on $\mathbb{S}^d$, by performing the path integral over $\sigma_{1}$ and $\sigma_2$. $G^{\mathbb{S}^d}_i$ is the two-point function of $O_i$ on the sphere of radius $R$
\begin{equation}
G_i^{\mathbb{S}^d}(x,y) =  \frac{1}{(R \, s(x,y))^{2 \Delta_i}}~,
\end{equation}
where $s$ is the distance between the two points $x,y$ induced by the round metric $g$ on the sphere of radius 1. Moreover $1^{\mathbb{S}^d}_{x,y} = \frac{1}{R^d \sqrt{g(x)}} \delta^d(x-y)$ and $\star$ is the product
\begin{equation}\label{sphereproduct}
(G^{\mathbb{S}^d}_1\star G^{\mathbb{S}^d}_2)(x,y) = \int_{\mathbb{S}^d}d^d z R^d\sqrt{g(z)} G^{\mathbb{S}^d}_1(x,z)G^{\mathbb{S}^d}_2(z,y)~.
\end{equation}
Taking the logarithm of \eqref{pfsphere} we have
\begin{equation}
\tilde{F}_f - \tilde{F}_0 = - \sin\left(\frac{\pi d}{2}\right) \frac 12 \log \det (1 - (f R^{d-\Delta_1-\Delta_2})^2 s^{-2 \Delta_1} \star s^{-2 \Delta_2})~.
\end{equation}
We want to compute the difference between the values in the deep UV and in the deep IR. Those are obtained by taking $f R^{d-\Delta_1-\Delta_2}$ to be 0 or $\infty$, respectively. We obtain
\begin{align}\label{deltafF}
\delta_f \tilde{F} = \tilde{F}_f^{UV} - \tilde{F}_f^{IR} & = \sin\left(\frac{\pi d}{2}\right)\frac 12 \log \det (s^{-2 \Delta_1} \star s^{-2 \Delta_2}) \nonumber \\ & = \sin\left(\frac{\pi d}{2}\right)\frac 12 \left(\log \det (s^{-2 \Delta_1}) + \log \det(s^{-2 \Delta_2})\right) \\ & \equiv \delta \tilde{F}_{\Delta_1} + \delta \tilde{F}_{\Delta_2} ~. \nonumber
\end{align}
Comparing \eqref{deltafF} with eq. (3.4) in \cite{Giombi:2014xxa}, we see that this coincides with the variation of $\tilde{F}$ induced by the deformation $g_1 O_1^2 + g_2 O_2^2$. This agrees with the fact that the deformations $f O_1 O_2$ and $g_1 O_1^2 + g_2 O_2^2$ connect the same UV and IR fixed points. In \cite{Diaz:2007an, Giombi:2014xxa} the logarithm of the functional determinant was evaluated via an appropriate regularization of the infinite sum, and the end result shown to be positive whenever $\frac d2 -1 < \Delta_i <\frac d2$ . We refer to these papers for an explicit expression (see also \cite{Fei:2015oha}).

In the limit $\Delta_1 \to \frac{d}{2}-1$, the part of $\delta_f \tilde{F}$ that depends on $\Delta_1$ equals the value of $\tilde{F}$ for the CFT of a free scalar, and we have
\begin{equation}
\delta_f \tilde{F} = \tilde{F}_{\text{scalar}} +  \delta \tilde{F}_{\Delta_2}~,
\end{equation}
which is again a positive quantity if  $\frac d2 -1 < \Delta_2 <\frac d2 $, since  $ \tilde{F}_{\text{scalar}}  >0$ \cite{Giombi:2014xxa}. This equation reflects the fact that along the flow the free scalar and the primary single-trace operator of dimension $\Delta_2$ recombine, giving in the IR one primary single-trace operator of dimension $d-\Delta_2$. 

The upshot is then that the generalized F-theorem holds for the double-trace deformation \eqref{pertf}, and it does so also when multiplets recombine.

\bibliographystyle{plainnat}

\end{document}

%% file: macros.tex
\def\){\right)}
\def\({\left( }
\def\]{\right] }
\def\[{\left[ }
\newcommand{\la}{\langle}
\newcommand{\ra}{\rangle}

\def\NO{\nonumber}

\newcommand{\be}{\begin{equation}}
\newcommand{\ee}{\end{equation}}

\def\bea{\begin{eqnarray}}
\def\eea{\end{eqnarray}}

\def\bal#1\eal{\begin{align}#1\end{align}}

\def\bald{\begin{aligned}}
\def\eald{\end{aligned}}

\def\beqx{\begin{displaymath}}
\def\eeqx{\end{displaymath}}

\newcommand{\bmat}{\left(\begin{array}}
\newcommand{\emat}{\end{array}\right)}

\def\p{\pi}

\def\s{\sigma}

\def\D{\Delta}
\def\F{\Phi}
\def\G{\Gamma}

\def\bo{{\raise-.3ex\hbox{\large$\Box$}}}               % D'Alembertian
\def\face{{\raise.2ex\hbox{$\displaystyle \bigodot$}\mskip-2.2mu \llap {$\ddot
        \smile$}}}                                   % happy face
\def\>{\rangle}                                      %right angle
\def\<{\langle}                                      %left angle

\def\leftrightarrowfill{$\mathsurround=0pt \mathord\leftarrow \mkern-6mu
        \cleaders\hbox{$\mkern-2mu \mathord- \mkern-2mu$}\hfill
        \mkern-6mu \mathord\rightarrow$}        % <--> double differential
\def\dvec#1{\vbox{\ialign{##\crcr
        \leftrightarrowfill\crcr\noalign{\kern-1pt\nointerlineskip}
        $\hfil\displaystyle{#1}\hfil$\crcr}}}           % <--> accent
\def\-{\hphantom{-}}